# ThCr$_2$Si$_2$-type Ru-based superconductors LaRu$_2$M$_2$ (M = P and As): An *ab-initio* investigation

Md. Zahidur Rahaman[1]

*Department of Physics, Pabna University of Science and Technology,
Pabna-6600, Bangladesh*
*zahidur.physics@gmail.com*

Md. Atikur Rahman[2*]

*Department of Physics, Pabna University of Science and Technology,
Pabna-6600, Bangladesh*
*atik0707phy@gmail.com*
[*]Corresponding author.

## Abstract

ThCr$_2$Si$_2$-type Ru-based superconductors LaRu$_2$P$_2$ and LaRu$_2$As$_2$ with $T_c$ ~ 4 K and 7.8 K respectively have been reported very recently. Using the first principles method we investigate the detailed physical and superconducting properties including the structural properties, electronic properties, elastic properties and Thermodynamic properties of these superconductors. Our investigated lattice parameters accord well with the experimental result. The analysis of the electronic band structure and density of states (DOS) indicates the metallic nature of both the compounds emerges from the La and Ru and the study of chemical bonding implies that a mixture of covalent, ionic and metallic bonds exists in both the compounds. The study of the mechanical properties revels that LaRu$_2$P$_2$ is brittle in nature while LaRu$_2$As$_2$ demonstrates the ductile nature and both the compounds show anisotropic characteristics. Finally we determine the Debye temperature and detailed superconducting parameters indicating that both the compounds under study are phonon-mediated medium coupled BCS superconductors.

## 1. Introduction

Recently AM$_2$X$_2$ structured compounds (where, A is a lanthanide element or any alkaline earth element; M is any transition metal; X = P, Ge, Si or As) have gained huge interest due to their many attractive physical properties such as mixed valency, valence fluctuation, heavy fermion nature, superconductivity at high and low temperature etc. These transition metal compounds with AM$_2$X$_2$ type structure show very attractive and rich physics because of their close energies concerning the spin, charge and orbital dynamics [1, 2]. Ternary intermetallic compounds with more than two thousand representatives [3] basically derived from the BaAl$_4$ type structures are considered as one of the most crucial families of intermetallic. Among these representatives ThCr$_2$Si$_2$ type structure was first reported and described in 1965 by Ban and Sikirica [4]. In 1996 G. Just and P. Paufler represents a complete geometric examination of approximately six hundred compounds with ThCr$_2$Si$_2$ type structure [5]. Recently ThCr$_2$Si$_2$ structure type has gained great attention of researchers after discovering a new superconductor $(Ba_{0.6}K_{0.4})Fe_2As_2$ belongs to the "122" family of iron-arsenides with $ThCr_2Si_2$ type structure exhibits high transition temperature 38 K [6]. On the other side, Pt, Ni and Pd-based $ThCr_2Si_2$ type borocarbides [7-9] have been discovered in the recent years with the transition temperature up to 23 K which raises the hope to constitute a family of new high temperature superconductors.

In iron-based compounds, the ternary intermetallic (122-type compounds) with $ThCr_2Si_2$ type structure such as $AFe_2As_2$ (A = Sr, Ca, Ba, etc.) are free of oxygen with metallic nature [10-12]. These compounds have been extensively studied for deciphering their superconducting mechanism [13, 14]. Because of the presence of Ru in the same group of iron, similar superconductivity can be expected in Ru-based materials [15]. Among Ru-based superconductors, $Sr_2RuO_4$ is a familiar p-wave pairing superconductor with superconducting transition temperature of 0.93 K [16]. The ternary lanthanum ruthenium phosphide $LaRu_2P_2$ crystallizes in $ThCr_2Si_2$ type structure has found to be a superconductor with transition temperature of 4.0 K [17]. This superconductor has a same type of structure as that of parent phase $BaFe_2As_2$ [18]. Hence it is interesting to study the detailed physical properties of these materials to have a profound view about the superconducting mechanism. Some experimental studies have also been done to investigate the superconducting and electronic properties of $LaRu_2P_2$ [19-22]. The isotropic superconductivity in $LaRu_2P_2$ was investigated by Ying and his co-workers [20]. Hass-van Alphen oscillations of this compound was investigated by Moll and his co-workers [21] and later following this path Razzoli et al shows that the superconducting phase in $LaRu_2P_2$ and one in the 122-Fe pnictides is not the same [22]. The ternary lanthanum ruthenium arsenide $LaRu_2As_2$ with $ThCr_2Si_2$ type structure is synthesized by Qi Guo et al [15]. He and his co-workers reported the bulk superconductivity in this compound with $T_c \sim 7.8$ K. The magnetization measurements of $LaRu_2As_2$ indicate that it is a bulk type-II superconductor [15].

Though notable advancement has been made to study the superconducting and electronic properties of lanthanum ruthenium phosphide, there is no theoretical and experimental works available in the literature about the detailed mechanical and thermodynamic properties of this superconductor. On the other hand, there is no study available except the superconducting mechanism of $LaRu_2As_2$ superconductor. We therefore decide to investigate the detailed physical properties including structural, elastic, electronic, thermodynamic and superconducting properties of these newly reported ternary intermetallic superconductors $LaRu_2P_2$ and $LaRu_2As_2$. We have used the density functional theory based CASTEP computer program to investigate the detailed physical properties of these compounds to have a profound understanding about the detailed physical nature of these superconductors and then a comparison has been made among the physical characteristics of these superconductors.

## 2. *Ab-initio* Computation details

All the calculations were carried out using the density functional theory dependent CASTEP computer code by employing GGA which stands for generalized gradient approximation with the PBE (Perdew-Burke-Ernzerhof) exchange correlation function [23-27]. The P-$3s^2$ $3p^3$, Ru-$4s^2$ $4p^6$ $4d^7$ $5s^1$ and La-$5s^2$ $5p^6$ $5d^1$ $6s^2$ in case of $LaRu_2P_2$ superconductor and As-$4s^2$ $4p^3$, Ru-$4s^2$ $4p^6$ $4d^7$ $5s^1$ and La-$5s^2$ $5p^6$ $5d^1$ $6s^2$ in case of $LaRu_2As_2$ superconductor were considered as valence electrons for pseudo atomic calculation. The Monkhorst-Pack scheme [28] was used to construct the K-point sampling of Brillouin zone. To expand the wave functions 400 eV cut-off energy was employed for both the compounds with 10×8×6 grids (236 irreducible k-points) and 8×8×6 grids (104 irreducible k-points) in the primitive cell of $LaRu_2P_2$ and $LaRu_2As_2$ superconductors respectively. The optimized structure for both the compounds was determined by using the Brodyden-Fletcher-Goldfarb-Shanno (BFGS) scheme [29].

The stress-strain method was used to determine the elastic stiffness constants of $LaRu_2P_2$ and $LaRu_2As_2$ ternary intermetallics [30]. The convergence criteria were fixed to $2.0 \times 10^{-6}$ eV/atom, $2.0 \times 10^{-4}$ Å and 0.006 eV/ Å for energy, maximum ionic displacement and maximum ionic force respectively. The maximum strain amplitude was fixed to 0.003.

## 3. Results and discussion

### 3.1. Electronic Structure

The ternary intermetallic compounds $LaRu_2P_2$ and $LaRu_2As_2$ belong to the tetragonal crystallographic system with space group *I4/mmm* (139). Both the superconductors belong to $ThCr_2Si_2$-type crystal structure with the Wyckoff position 2*a* (0, 0, 0) for La, 4*d* (0, 0.5, 0.25) for Ru and 4*e* (0, 0, 0.3593) for P or As [17]. There are ten atoms per unit cell with two formula units and five atoms per primitive cell with one formula unit. For obtaining the fully relaxed structure of these two superconductors we optimize the geometry including the lattice parameters and atomic positions as a function of normal stress as illustrated in Fig.1. The evaluated equilibrium lattice parameters $a_0$ and $c_0$, unit cell volume $V_0$ and bulk modulus $B_0$ for $LaRu_2P_2$ and $LaRu_2As_2$ superconductors at 0 K are tabulated in Table 1 with the available experimental values. The calculated structural parameters in this present study accord well with the experimental values. This bears the reliability of the present DFT-based first-principles calculations. It can be noted from Table 1 that the replacement of P by As mostly influences the *a* values whereas a slight change is appeared in case of *c* values. A slight change in bulk modulus is appeared when P is substituted by As. More theoretical work is required to comprehend this intriguing discrepancy.

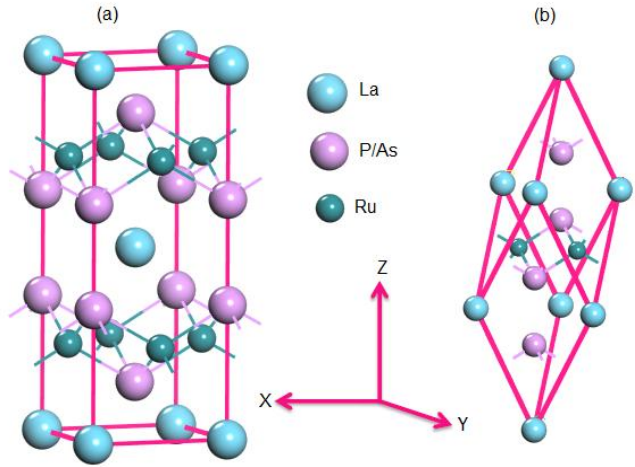

**Fig. 1.** The crystal structures of $LaRu_2M_2$ (M = P and As) (a) conventional unit cell and (b) primitive cell.

**Table 1.** Unit cell dimensions of ThCr$_2$Si$_2$-type LaRu$_2$P$_2$ and LaRu$_2$As$_2$ superconductors.

| Properties | LaRu$_2$P$_2$ | | LaRu$_2$As$_2$ | |
| --- | --- | --- | --- | --- |
| | This study | Expt. [17] | This study | Expt. [17] |
| $a_0$ (Å) | 4.035 | 4.031 | 4.209 | 4.180 |
| $c_0$ (Å) | 11.019 | 10.675 | 10.921 | 10.623 |
| $c_0/a_0$ | 2.730 | 2.648 | 2.594 | 2.541 |
| $V_0$ (Å$^3$) | 179.40 | 173.45 | 193.47 | 185.60 |
| $B_0$ (GPa) | 121.22 | - | 113.15 | - |

*3.2. Mechanical Properties*

The elastic constants provide crucial information about the nature of force in a solid. It provides deep insight into the mechanical and dynamical behavior of different crystals. The study of elastic constants provides evident idea about different materials properties including anisotropy, ductility, stability, brittleness and stiffness of materials [31]. In this section we have discussed about the detailed mechanical behavior of LaRu$_2$P$_2$ and LaRu$_2$As$_2$ superconductors. The elastic constants of both the compounds were obtained from a linear fit of the evaluated stress-strain function according to Hook's law [32]. The tetragonal phase of crystal has six independent elastic constants $C_{11}$, $C_{12}$, $C_{13}$, $C_{33}$, $C_{44}$ and $C_{66}$. The evaluated elastic constants for both the superconductors are tabulated in Table 2. The mechanical stability criteria for the tetragonal phase are given as follows [33],

$$\left.\begin{array}{l} C_{11} > 0,\ C_{44} > 0,\ C_{33} > 0,\ C_{66} > 0 \\ C_{11} + C_{33} - 2C_{13} > 0,\ C_{11} - C_{12} > 0 \\ 2(C_{11} + C_{12}) + C_{33} + 4C_{13} > 0 \end{array}\right\} \quad (1)$$

Our evaluated elastic constants for both the compounds (Table 2) satisfy the above stability criteria showing that LaRu$_2$P$_2$ and LaRu$_2$As$_2$ superconductors are stable in nature. From Table 2 it can be seen that $C_{11} > C_{33}$ for both the compounds showing that the atomic bonding between (100) and (001) planes does not have the same strength [34] and the incompressibility along [100] direction is stronger than [001] direction [35]. Again $C_{66}$ is more larger than $C_{44}$ indicating that the [100](001) shear is easier than the [100](010) shear for both the intermetallics [35]. Most of the elastic constants of LaRu$_2$P$_2$ are larger compared with LaRu$_2$As$_2$ indicating that the shear resistance and the incompressibility of LaRu$_2$P$_2$ compound are stronger than LaRu$_2$As$_2$.

**Table 2.** The calculated elastic constants $C_{ij}$ (in GPa) of LaRu$_2$P$_2$ and LaRu$_2$As$_2$ superconductors.

| Compounds | Elastic constants | | | | | |
|---|---|---|---|---|---|---|
| | $C_{11}$ | $C_{12}$ | $C_{13}$ | $C_{33}$ | $C_{44}$ | $C_{66}$ |
| LaRu$_2$P$_2$ | 277.52 | 104.54 | 72.26 | 116.39 | 50.64 | 132.75 |
| LaRu$_2$As$_2$ | 242.95 | 88.49 | 82.96 | 95.72 | 52.60 | 94.44 |

By using the Voigt-Reuss-Hill (VRH) averaging scheme [36] we have determined the most important mechanical properties such as the Young's modulus $E$, bulk modulus $B$, shear modulus $G$ and Poisson's ratio $v$. For tetragonal structure, the Voigt and Reuss approximation for the bulk and shear moduli are as follows:

$$B_V = \frac{2C_{11} + 2C_{12} + C_{33} + 4C_{13}}{9} \quad (2)$$

$$B_R = \frac{C^2}{M} \quad (3)$$

$$G_V = \frac{M + 3C_{11} - 3C_{12} + 12C_{44} + 6C_{66}}{30} \quad (4)$$

$$G_R = \frac{15}{\left[\frac{18B_V}{C^2} + \frac{6}{(C_{11} - C_{12})} + \frac{6}{C_{44}} + \frac{3}{C_{66}}\right]} \quad (5)$$

Where,

$$M = C_{11} + C_{12} + 2C_{33} - 4C_{13}$$

And

$$C^2 = (C_{11} + C_{12})C_{33} - 2C_{13}^2$$

The Hill took an arithmetic mean of $B$ and $G$ given below,

$$B = \frac{1}{2}(B_R + B_v) \quad (6)$$

$$G = \frac{1}{2}(G_v + G_R) \quad (7)$$

Now we can calculate the Young's modulus ($E$), and Poisson's ratio ($v$) by using following relations,

$$E = \frac{9GB}{3B + G} \quad (8)$$

$$v = \frac{3B - 2G}{2(3B + G)} \quad (9)$$

The calculated polycrystalline elastic parameters by using Eq. 2 to Eq. 9 are presented in Table 3. One can see from Table 3 that the evaluated bulk modulus of both the compounds is comparatively large (> 100 GPa), hence these compounds can be classified as relatively hard materials. The lanthanum ruthenium phosphide has a strong resistance to volume change under applied force than lanthanum ruthenium arsenide since the value of $B$ in case of $LaRu_2P_2$ is larger than that of $LaRu_2As_2$ [37]. Compared with $LaRu_2P_2$, the evaluated shear modulus of $LaRu_2As_2$ is smaller indicating weaker covalent bond and lower shear resistance in $LaRu_2As_2$ [38]. Again the obtained young's modulus of $LaRu_2P_2$ is larger than that of $LaRu_2As_2$ (Table 3) demonstrating that $LaRu_2P_2$ is much stiffer than $LaRu_2As_2$ [39].

The ratio between bulk and shear modulus ($B/G$) is known as Pugh's ratio which is used to predict the brittleness and ductility of material [39]. $B/G < 1.75$ demonstrates the brittle manner of a compound otherwise the material will be ductile. From Table 3 one can notice that $LaRu_2P_2$ is brittle in nature while $LaRu_2As_2$ demonstrates the ductile nature. The Poisson's ratio is used to comprehend the nature of bonding force in crystal [40]. According to the value (Table 3) of Poisson's ratio $LaRu_2P_2$ superconductor can be classified as an ionic crystal since for ionic crystal the value of Poisson's ratio is 0.25 [41]. Whereas the force exists in $LaRu_2As_2$ superconductor is central since $0.25 < \nu < 0.50$ indicates the central force in a crystal [41].

The anisotropic characteristics of a crystal can be estimated by the following equation [42],

$$A^U = \frac{5G_V}{G_R} + \frac{B_V}{B_R} - 6 \qquad (10)$$

$A^U$ is zero for completely isotropic materials otherwise the material will be anisotropic. From Table 3 we see that both compounds show large anisotropic characteristics whereas $LaRu_2As_2$ shows large anisotropy than $LaRu_2P_2$. Again for comparison we determine the anisotropy indexes of shear and bulk moduli ($A_G$ and $A_B$) suggested by Chung and Buessen [43] as follows,

$$A_G = \frac{(G_V - G_R)}{(G_V + G_R)} \qquad (11)$$

$$A_B = \frac{(B_V - B_R)}{(B_V + B_R)} \qquad (12)$$

Where, $A_B = A_G = 1$ represents the maximum anisotropy and $A_B = A_G = 0$ shows completely isotropic characteristics. For tetragonal $LaRu_2P_2$, $A_B = 0.10$ and $A_G = 0.10$ and for tetragonal $LaRu_2As_2$ superconductor $A_B = 0.12$ and $A_G = 0.11$ demonstrating the anisotropic characteristics of both the superconductors. The hardness of materials can be predicted by using a new theoretical model suggested by Chen et al. as follows [43],

$$H_V = 2(K^2 G)^{0.585} - 3 \qquad (13)$$

According to the values listed in Table 3, $LaRu_2P_2$ can be classified as relatively hard material whereas $LaRu_2As_2$ represents relatively soft characteristics.

**Table 3.** Calculated polycrysttalline bulk modulus *B* (GPa), shear modulus *G* (GPa), Young's modulus *E* (GPa), *B/G* values, Poisson's ratio *v*, elastic anisotropy $A^U$ and Vickers hardness $H_v$ (GPa) of $LaRu_2P_2$ and $LaRu_2As_2$ superconductors.

| | Polycrystalline elastic properties | | | | | | |
|---|---|---|---|---|---|---|---|
| Compounds | *B* | *G* | *E* | *B/G* | *v* | $A^U$ | $H_v$ |
| $LaRu_2P_2$ | 117.19 | 68.15 | 171.25 | 1.71 | 0.25 | 1.35 | 9.53 |
| $LaRu_2As_2$ | 107.58 | 55.37 | 141.78 | 1.94 | 0.28 | 1.58 | 6.62 |

*3.3. Electronic properties and chemical bonding*

To gain deep insights into the electronic properties and chemical bonding in $LaRu_2P_2$ and $LaRu_2As_2$ superconductors, the electronic band structure, partial density of states (PDOS) of La, Ru, P and As, total density of states (TDOS), total charge density and charge density difference for both the superconductors have been studied and discussed. The electronic band structures for both the ternary intermetallic compounds have been illustrated in Fig.2, according to which both the compounds are metallic since no band gap is appeared at $E_f$. The metallic nature of $LaRu_2P_2$ and $LaRu_2As_2$ implies that these two compounds might be superconductor [31].

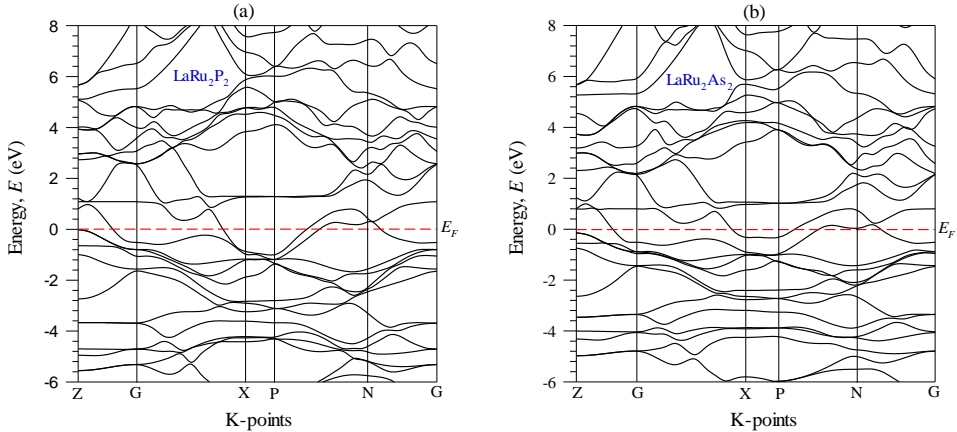

**Fig. 2.** Electronic band structure of (a) $LaRu_2P_2$ and (b) $LaRu_2As_2$ superconductors along high symmetry direction in the Brillouin zones.

The partial and total density of states of $LaRu_2P_2$ and $LaRu_2As_2$ compounds are plotted in Fig.3. From -6 eV to Fermi level ($E_f$) the dominant feature of hybridization is observed for La-5d, Ru-4d and P-3p in case of $LaRu_2P_2$ and La-5d, Ru-4d and As-4p in case of $LaRu_2As_2$ superconductor. The valance bands mostly originate from Ru-4d orbital for both the compounds. However at Fermi level domination of La-5d and Ru-4d orbital is observed for both the superconductors. The conduction bands mostly consist of La-5d, Ru-4d and P-3p in case of $LaRu_2P_2$ and La-5d, Ru-4d and As-4p in case of $LaRu_2As_2$ superconductor where the contribution of La-5d state is dominant for both the compounds. The metallic nature of these superconductors mainly originates from La and Ru metals with some contribution of P and As. Coincidence between Ru-4d and P-3p states implies the covalent nature of P-Ru bonds in case of LaRu$_2$P$_2$. Similarly the coincidence between Ru-4d and As-4p states implies the covalent feature of As-Ru bonds in case of LaRu$_2$As$_2$ superconductor. These results agree well with the experimentally observed results [17]. The evaluated DOS at Fermi level is 2.60 states eV$^{-1}$ fu$^{-1}$ and 2.72 states eV$^{-1}$ fu$^{-1}$ respectively for LaRu$_2$P$_2$ and LaRu$_2$As$_2$ superconductors.

**Table 4.** Mulliken atomic populations of LaRu$_2$P$_2$ and LaRu$_2$As$_2$ superconductors.

|  | Species | s | p | d | Total | Charge | Bond | Population | $f_h$ | Lengths (Å) |
|---|---|---|---|---|---|---|---|---|---|---|
| **LaRu$_2$P$_2$** | P | 1.58 | 3.33 | 0.00 | 4.91 | 0.09 | P-Ru | 0.93 | 0.072 | 2.3286 |
|  | Ru | 2.52 | 6.93 | 7.35 | 16.81 | -0.81 | Ru-Ru | -0.42 | … | 2.8538 |
|  | La | 1.77 | 5.98 | 1.81 | 9.56 | 1.44 |  |  |  |  |
| **LaRu$_2$As$_2$** | As | 1.56 | 3.30 | 0.00 | 4.86 | 0.14 | As-Ru | -3.76 | … | 2.4369 |
|  | Ru | 2.53 | 6.89 | 7.35 | 16.77 | -0.77 | Ru-Ru | 1.60 | 0.312 | 2.9763 |
|  | La | 1.94 | 6.02 | 1.78 | 9.74 | 1.26 |  |  |  |  |

To gain more comprehensive insights into LaRu$_2$M$_2$ (M = P and As) superconductors, Mulliken atomic populations [44] of these two superconductors have been investigated and analyzed which is particularly useful for the information about the chemical bonding nature of materials. A low value of the bond population indicates the ionic character (For a perfectly ionic interaction the value of the bond population is zero) whereas a high value expresses increasing level of covalency [45]. The calculated bond populations of LaRu$_2$M$_2$ are tabulated in Table 4. From Table 4 we notice that in case of LaRu$_2$P$_2$ superconductor P and La atoms carry the positive charges while only Ru atoms carry the negative charges indicating the transfer of charge from P and La to Ru atoms. Similarly for LaRu$_2$As$_2$ superconductor, charges transfer from As and La to Ru atoms. The transferred charge from P to Ru and La to Ru are equal to 0.81$e$ in case of LaRu$_2$P$_2$ and from As to Ru and La to Ru in case of LaRu$_2$As$_2$ superconductor are equal to 0.77$e$ suggesting an effective valence state of P$^{0.09}$La$^{1.44}$Ru$^{-0.81}$ and As$^{0.14}$La$^{1.26}$Ru$^{-0.77}$ for LaRu$_2$P$_2$ and LaRu$_2$As$_2$ superconductors respectively. We see from Table 4 that the bond population of P-Ru bond is 0.93 which is positive and greater than zero indicating the strong covalent interaction of P and Ru atoms in LaRu$_2$P$_2$. On the other hand the bond population of Ru-Ru bond is -0.42 indicating the ionic character of this bond in LaRu$_2$P$_2$. But in case of LaRu$_2$As$_2$ we observe totally opposite interaction. From Table 3 it is evident that the interaction among As and Ru atoms is strongly ionic whereas the nature of Ru-Ru bond is highly covalent in LaRu$_2$As$_2$.

For further understanding the bonding characteristics we have calculated the ionicity of a bond by using the following equation [46],

$$f_h = 1 - e^{-|P_c - P|/P} \qquad (14)$$

Where, $P$ is the bond overlap population and $P_c$ is the bond overlap population in a pure covalent crystal. For pure covalent crystal, the value of $P_c$ is 1. Unit value of $f_h$ represents a pure ionic bond whereas the zero value indicates the pure covalent bond. From Table 4 it is clear that in LaRu$_2$P$_2$ the P-Ru bonds show high level of covalency agreeable with the result having from DOS analysis and Mulliken atomic population analysis. In LaRu$_2$As$_2$ the Ru-Ru bonds show high level of covalency and low level of ionicity which is also agreeable with the result having from Mulliken atomic population analysis.

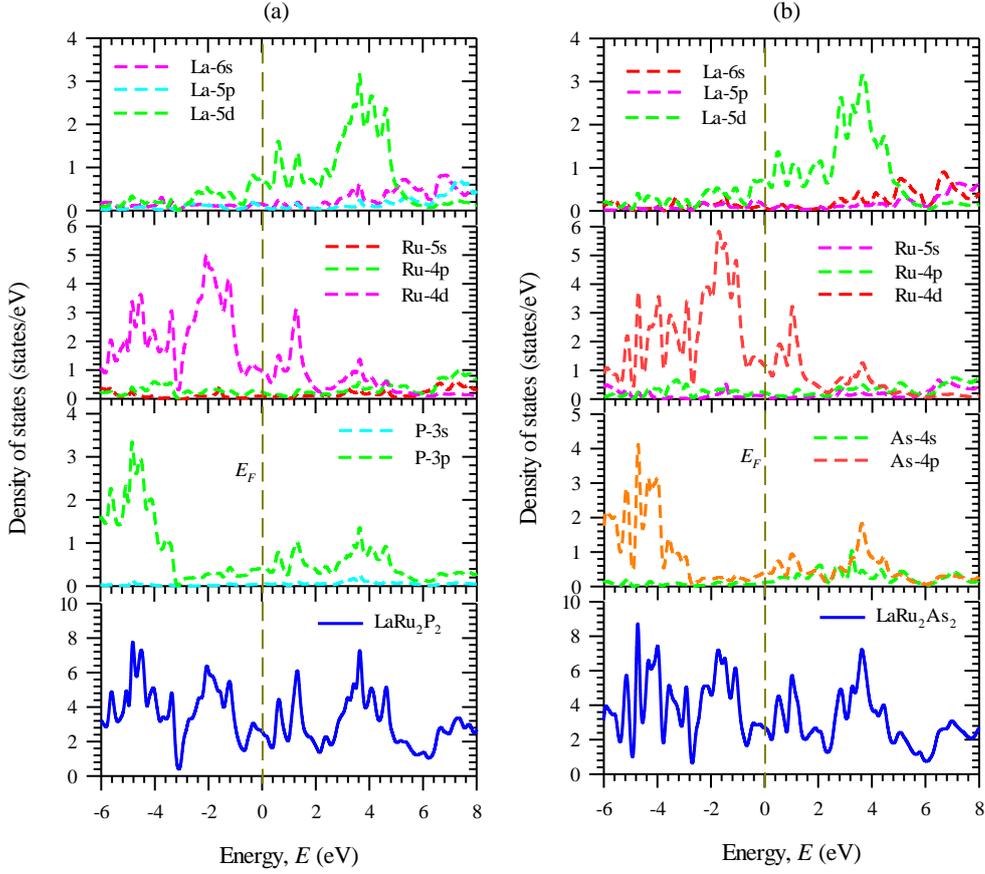

**Fig. 3.** The partial and total density of states of (a) LaRu$_2$P$_2$ and (b) LaRu$_2$As$_2$ superconductors.

The detailed investigation of total charge density and charge density difference of materials are particularly useful for analyzing chemical bonding and charge transfer within a compound. For further understanding the bonding characteristics of these two superconductors precisely we study the charge density difference (in the unit of e/Å$^3$) and total charge density in the direction of (200) crystallographic plane as shown in Fig. 4 and Fig. 5 for LaRu$_2$P$_2$ and LaRu$_2$As$_2$ superconductors respectively.

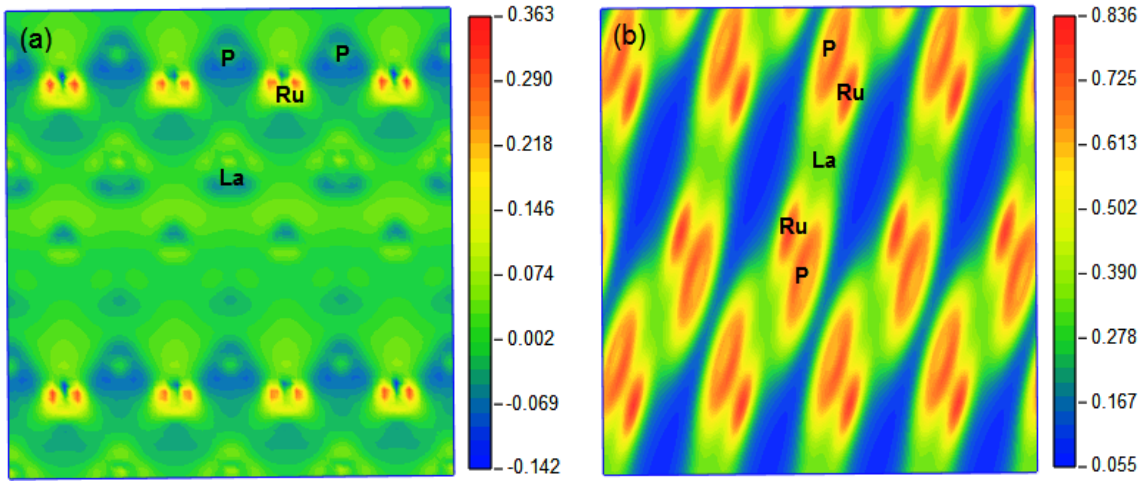

**Fig. 4.** (a) The charge density difference and (b) the total charge density of LaRu$_2$P$_2$ superconductor for (200) plane.

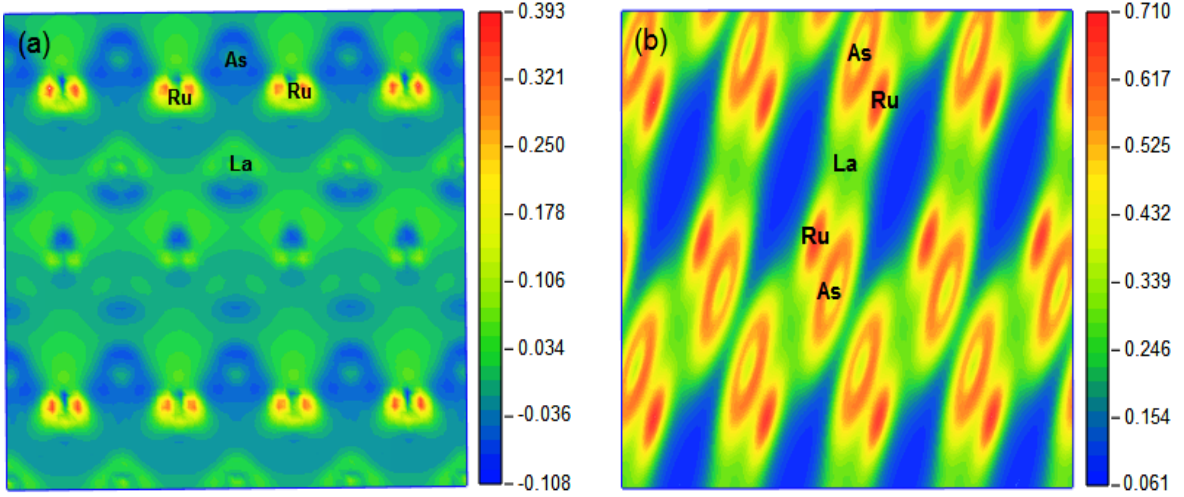

**Fig. 5.** (a) The charge density difference and (b) the total charge density of LaRu$_2$As$_2$ superconductor for (200) plane.

A scale is shown with each contour plots at the right which represents the intensity of charge density. The high and light density of charge (electron) is represented by red and blue color respectively.

Exchange of electrons in space is represented by the charge density difference map [47]. In Fig. 4(a) and Fig. 5(a) the gain and loss of electron is indicated in red and blue color respectively. It is evident from Fig. 4(a) that P and La atoms lose electrons while the Ru atoms gain electrons indicating the transfer of charge from P and La atoms to Ru atoms. These show some ionic interaction of Ru with P and La atoms in LaRu$_2$P$_2$. Similarly it is clear from Fig. 5(a) that the exchange of electrons occurs from La and As to Ru atoms indicating the existence of ionic interaction among As-Ru and La-Ru bonds. For both the cases we observe strong ionic interaction of Ru atoms with P and As while a very weak interaction is occurred among La and Ru atoms. Large interatomic distance between La and Ru compared to that of Ru with P and As may be the reason for this kind of interaction. We also observe strong ionic interaction between P-Ru bonds compared to that of As-Ru bonds.

In Fig. 4(b) we observe a clear overlapping of charge distribution between nearest P and Ru atoms indicating the covalent nature of P-Ru bonds. The similar interaction is observed between As-Ru bonds from Fig. 5(b). On the other hand a spherical like charge distribution is observed around Ru atoms and evidently no overlapping of electron distribution occur indicating the ionic nature of Ru-Ru bonds for both the superconductors. The ionic characteristics is a result of the metallic characteristics [48] indicating the metallic nature of Ru-Ru bonds.

Therefore, we can conclude that both the compounds may be described as a mixture of ionic, covalent and metallic bonds.

### 3.4. Thermodynamic and superconducting properties

The temperature which is related to the highest normal mode of vibration of a crystal is generally referred to as Debye temperature. It gives deep insight into many crucial thermodynamic properties of materials such as melting point, specific heat, thermal expansion etc. However various estimations can be used to calculate the Debye temperature of a material. In this present study, we have used the evaluated elastic constants data to calculate the Debye temperature ($\Theta_D$) of LaRu$_2$P$_2$ and LaRu$_2$As$_2$ superconductors. In this way $\Theta_D$ can be calculated by using the following equation [49]:

$$\theta_D = \frac{h}{k_B}\left(\frac{3N}{4\pi V}\right)^{\frac{1}{3}} \times v_m \qquad (15)$$

Where, $V_m$ is defined as the average sound velocity which can be obtained,

$$v_m = \left[\frac{1}{3}\left(\frac{2}{v_t{}^3} + \frac{1}{v_l{}^3}\right)\right]^{-\frac{1}{3}} \quad (16)$$

Where, $V_t$ is defined as the transverse wave velocity and $V_l$ stands for longitudinal wave velocity which are given by,

$$v_l = \left(\frac{3B + 4G}{3\rho}\right)^{\frac{1}{2}} \quad (17)$$

And

$$v_t = \left(\frac{G}{\rho}\right)^{\frac{1}{2}} \quad (18)$$

The evaluated values of $V_t$, $V_l$ and $V_m$ of LaRu$_2$P$_2$ and LaRu$_2$As$_2$ are illustrated in Table 5. Substituting these values in Eq. (15) the obtained Debye temperature ($\Theta_D$) of LaRu$_2$P$_2$ and LaRu$_2$As$_2$ superconductors is 392.51 K and 317.56K respectively. Due to the absence of any experimental and theoretical data it is not possible to determine the magnitude of error.

**Table 5.** The evaluated density $\rho$ (in gm/cm$^3$), transverse ($V_t$), longitudinal ($V_l$), and average sound velocity $V_m$ (m/s) and Debye temperature $\Theta_D$ (K) of LaRu$_2$P$_2$ and LaRu$_2$As$_2$ superconductors.

| Compounds | $\rho$ | $V_t$ | $V_l$ | $V_m$ | $\Theta_D$ |
|---|---|---|---|---|---|
| LaRu$_2$P$_2$ | 7.43 | 3028.57 | 9165.51 | 3446.19 | 392.51 |
| LaRu$_2$As$_2$ | 8.39 | 2568.95 | 4649.92 | 2862.15 | 317.56 |

The transition temperature ($T_c$) of a superconductor can be calculated theoretically by using McMillan equation [50] as follows,

$$T_c = \frac{\theta_D}{1.45} e^{-\frac{1.04\,(1+\lambda)}{\lambda - \mu^*(1+0.62\lambda)}} \quad (19)$$

Where, $\theta_D$ stands for Debye temperature, $\mu^*$ is defined as the coulomb pseudo potential and $\lambda$ is denoted as the electron-phonon coupling constant. The value of $\lambda$ can be estimated as follows [51],

$$\lambda = \frac{\gamma_{exp}}{\gamma_{bs}} - 1 \quad (20)$$

Where, $\gamma_{exp}$ and $\gamma_{bs}$ is defined as the experimental electronic specific heat coefficient and theoretical electronic specific heat coefficient respectively. The theoretical electronic specific heat coefficient $\gamma_{bs}$ is given by [51],

$$\gamma_{bs} = \frac{\pi^2 K_B^2 N(E_F)}{3} \quad (21)$$

Substituting the value of $N(E_F)$ getting from the DOS analysis of this present study for LaRu$_2$P$_2$ and LaRu$_2$As$_2$ superconductors in Eq. 21 we evaluate the value of $\gamma_{bs}$ listed in Table 6. Then using Eq. 20 the electron-phonon coupling constant $\lambda$ has been evaluated as shown in Table 6. Finally putting the value of the Debye temperature in Eq. 19 the superconducting critical temperature of LaRu$_2$P$_2$ compound has been determined (Table 6). In Eq. 19 the Coulomb pseudo potential $\mu^*$ is calculated as [52],

$$\mu^* = 0.26 \frac{N(E_F)}{1 + N(E_F)} \quad (22)$$

Since no experimental value of electronic specific heat coefficient $\gamma_{exp}$ is available in case of LaRu$_2$As$_2$ compound, it is not possible to evaluate $T_c$ following the above way for this compound. So by using the following equation [50] we have directly evaluated the value of $\lambda$.

$$\lambda = \frac{1.04 + \mu^* \ln\left(\frac{\theta_D}{1.45 T_c}\right)}{(1 - 0.62\mu^*) \ln\left(\frac{\theta_D}{1.45 T_c}\right) - 1.04} \quad (23)$$

Here, we use the value of $T_c$ as 7.8 K (getting from experiment) for LaRu$_2$As$_2$ compound. We have also evaluated the value of electronic specific heat coefficient $\gamma$ as $\gamma = \gamma_{bs}(1+\lambda)$ where the phonon enhancement factor $(1+\lambda)$ is renormalized compared to the band structure value $\gamma_{bs}$ [54]. All these evaluated superconducting parameters for both the superconductors are tabulated in Table 6 with the available experimental and other theoretical values.

**Table 6.** Calculated density of states at Fermi level $N(E_F)$ (states eV$^{-1}$ fu$^{-1}$), specific heat coefficient $\gamma_{bs}$ (calculated by using band structure data) and $\gamma$ (mJ/ K$^2$ mol), electron-phonon coupling constant $\lambda$, the coulomb pseudo potential $\mu^*$ and transition temperature $T_c$ (K) of LaRu$_2$P$_2$ and LaRu$_2$As$_2$ ternary intermetallic compounds.

| Compounds | | $N(E_F)$ | $\gamma_{bs}$ | $\gamma$ | $\lambda$ | $\mu^*$ | $T_c$ |
|---|---|---|---|---|---|---|---|
| **LaRu$_2$P$_2$** | This Cal. | 2.60 | 6.12 | 11.44 | 0.87 | 0.18 | 10.18 |
| | Exp. [49] | - | - | - | 0.98 | - | 4.0 |
| | [53] | - | - | 11.50 | 0.98 | - | - |
| | Other Cal.[54] | 2.38 | - | 10.35 | 0.85 | 0.18 | 3.74 |
| | [53] | 2.46 | - | - | - | - | - |
| **LaRu$_2$As$_2$** | This Cal. | 2.72 | 6.40 | 12.03 | 0.88 | 0.19 | - |
| | Exp. [15] | - | - | - | - | - | 7.8 |
| | Other Cal. | - | - | - | - | - | - |

From Table 6 we notice that the calculated value of $\gamma$ and $\mu^*$ shows good concordance with the experimentally evaluated value in case of LaRu$_2$P$_2$ superconductor. We also note that the evaluated value of $\lambda$ is slightly lower than the experimentally evaluated value but shows relatively good agreement with the experimental value compared with the other theoretical values. However the calculated value of $T_c$ in case of LaRu$_2$P$_2$ superconductor is 10.18 K which shows conflicts with the experimental value 4 K. Since all the superconducting parameters which we have evaluated show good agreement with the experimental value, perhaps the reason for this confliction lies within the value of the Debye temperature. Since the Debye temperature is proportional to the value of $T_c$, the experimentally calculated value of $\Theta_D$ must be lower than that of our calculated value 392.51 K. Due to the lacking of available experimental data about LaRu$_2$As$_2$ superconductor, comparison of our evaluated data for LaRu$_2$As$_2$ compound is not possible. However the calculated value of electron-phonon coupling constant $\lambda$ suggests that both the compounds under study are phonon-mediated medium coupled Bardeen-Copper-Schrieffer (BCS) superconductors. From the DOS diagram as shown in Fig. 3 we notice that most of the contribution near Fermi level comes from Ru-4d states for both the compounds and hence the vibration related these states make large contribution to $\lambda$ [54]. Since the contribution of Ru-4d states is large than the others states in both the superconductors hence the contribution of 4d states electrons exhibits the possibility of superconductivity in LaRu$_2$P$_2$ and LaRu$_2$As$_2$ ternary intermetallics.

## 4. Conclusions

In summary, the detailed physical properties including structural, electronic, bonding, mechanical, thermodynamic and superconducting properties of two ternary intermetallic compounds LaRu$_2$P$_2$ and LaRu$_2$As$_2$ have been investigated via *ab-initio* method based on the density functional theory. The optimized lattice parameters demonstrate slight variation from experimental values. The study of band structure and DOS revels the metallic nature of both the compounds emerges mainly from the La and Ru atoms and the study of chemical bonding implies that a mixture of covalent, ionic and metallic bonds exist in both the compounds. The study of mechanical properties revels that both the compounds are mechanically stable and demonstrates anisotropic characteristics. The evaluated *B/G* ratio shows that LaRu$_2$P$_2$ is brittle in nature while LaRu$_2$As$_2$ demonstrates the ductile nature. The Vickers hardness revels that LaRu$_2$P$_2$ can be classified as relatively hard material whereas LaRu$_2$As$_2$ represents relatively soft characteristics. The Debye temperature of LaRu$_2$P$_2$ and LaRu$_2$As$_2$ superconductors are calculated as 392.51 K and 317.56 K respectively. Moreover, the superconducting transition temperature $T_c$ of LaRu$_2$P$_2$ is calculated to be 10.18 K which is larger than the experimental value 4 k. Due to the lacking of available experimental value of electronic specific heat coefficient of LaRu$_2$As$_2$ we are not being able to evaluate $T_c$ of LaRu$_2$As$_2$ compound. Using the experimental $T_c$ we directly evaluate the value of electron-phonon coupling constant of LaRu$_2$As$_2$ as 0.88. However the calculated value of electron-phonon coupling constant $\lambda$ suggests that both the compounds under study are phonon-mediated medium coupled Bardeen-Copper-Schrieffer (BCS) superconductors whereas the contribution of 4d states electrons exhibits the possibility of superconductivity in LaRu$_2$P$_2$ and LaRu$_2$As$_2$ ternary intermetallics.

## Acknowledgements

I (1$^{st}$ author) would like to thank Assistant Prof. Md. Atikur Rahman (Sir) for providing me the best support during the research period. I would also like to thank all of the honorable teachers of Physics Department, Pabna University of Science and Technology, Bangladesh, for their encouraging speech.

## Author Contributions

Md.Z. Rahaman conceived the idea and initiated the project. He also performed the first principle calculations and prepared all figures. He also analyzed the data and wrote the manuscript. Md.A. Rahman supervised the project.